\documentclass[3p,times,twocolumn]{elsarticle}

\usepackage{ecrc}
\usepackage{color}
\usepackage{graphicx}
\usepackage[utf8]{inputenc}
\usepackage{amsmath,amsfonts,amssymb,amsbsy}
\usepackage{multirow}
\usepackage{macros_alpha,defs}

\def\SF{Schr\"odinger functional}
\def\GF{Gradient Flow}
\def\PT{perturbation theory}

\volume{00}

\firstpage{1}

\journalname{Nuclear and Particle Physics Proceedings}

\runauth{}

\jid{nppp}

\jnltitlelogo{Nuclear and Particle Physics Proceedings}

\usepackage{amssymb}

\usepackage[figuresright]{rotating}

\begin{document}

\begin{frontmatter}



\author[desy,bnl]{Mattia~Bruno}
\author[desy]{Mattia~Dalla~Brida}
\author[uam,cern]{Patrick~Fritzsch} %
\author[hu,wup]{Tomasz~Korzec}
\author[cern]{Alberto~Ramos} 
\author[desy]{Stefan~Schaefer}
\author[desy]{Hubert~Simma}
\author[dublin]{Stefan~Sint}%
\author[desy,hu]{Rainer~Sommer}
%
%
\address[desy]{John von Neumann Institute for Computing (NIC), DESY, Platanenallee~6, 15738 Zeuthen, Germany}
\address[bnl]{Physics Department, Brookhaven National Laboratory, Upton, NY 11973, USA}
\address[uam]{Instituto de F\'{\i}sica Te{\'o}rica UAM/CSIC, Universidad Aut{\'o}noma de Madrid,\\
C/ Nicol{\'a}s Cabrera 13-15, Cantoblanco, Madrid 28049, Spain}
\address[wup]{Department of Physics, Bergische Universit\"at Wuppertal, Gau\ss str. 20,
42119 Wuppertal, Germany}
\address[cern]{CERN, Theory Division, Geneva, Switzerland}
\address[dublin]{School of Mathematics, Trinity College Dublin, Dublin
  2, Ireland}
\address[hu]{Institut~f\"ur~Physik, Humboldt-Universit\"at~zu~Berlin, Newtonstr.~15, 12489~Berlin, Germany}

\ead{Rainer.Sommer@desy.de}


\dochead{\small \tt DESY 16-214 \hfill CERN-TH-2016-236
    \hfill IFT-UAM/CSIC-16-119 \hfill WUB/16-10}

\title{The determination of $\alpha_s$ by the ALPHA collaboration
        \\[1.5ex]
        \includegraphics[width=2.0cm]{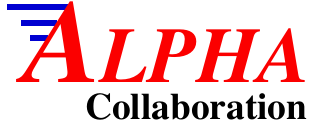}
        \vspace*{-3mm}
        }


\author{}

\address{}

\begin{abstract}
We review the ALPHA collaboration strategy for obtaining the 
QCD coupling at high scale. In the three-flavor 
effective theory it avoids the use of perturbation theory 
at $\alpha \grtsim 0.2$ and at the same time has the physical scales
small compared to the cutoff $1/a$ in all stages of the computation.
The result $\Lambda_\msbar^{(3)}=332(14)\,\MeV$ 
is translated to $\alpha_\msbar(m_Z)=0.1179(10)(2)$ by use of (high order) perturbative relations between the effective theory couplings at the 
charm and beauty quark ``thresholds''.
The error of this perturbative step is discussed and estimated as $0.0002$. 
\end{abstract}

\begin{keyword}
QCD \sep Perturbation theory \sep Strong coupling constant
\sep $\Lambda$-parameter \sep Lattice QCD

\end{keyword}

\end{frontmatter}


\section{Introduction }
\label{s:intro}
This  talk discusses the non-perturbative determination of $\alpha_s$,
using lattice QCD as the non-perturbative definition of the theory and
for its evaluation. One wants to relate
$\alpha_s(\mu)=\gbar^2_s(\mu)/(4\pi)$ (conventionally at $\mu=m_Z$ and in the $s=\msbar$ scheme)
to experimental observables with negligible truncation errors from the use of \PT\ (PT)
at intermediate scales. 
This is a very relevant task, since $\alpha_\msbar(\mu)$ enters
many important theory predictions, whether in LHC- or in flavor-physics.
But it seems that it is sometimes overlooked that it is also a 
true challenge to achieve a good systematic precision. Both the Particle Data Group \cite{Agashe:2014kda} 
and the Flavour Lattice Averaging Group \cite{Aoki:2016frl} 
are therefore not just taking weighted averages of the individual
determinations to arrive at their world averages. 

We start with a note on definitions of the QCD coupling and $\Lambda$-parameters which is needed
in order to understand what can be said non-perturbatively and what is
intrinsically perturbative. 

The standard is to use the $\msbar$ renormalization scheme for QCD.
Order by order in the coupling it defines
the relation between the bare coupling and the renormalized 
$\msbar$ coupling. There is no general definition of this relation
beyond this series, i.e. beyond PT. Therefore,
it is also hard to make firm statements about non-perturbative 
``contributions'' or ``corrections''. 
However, it is not hard to get around this conceptual and
practical problem.
One may start from some short-distance QCD
observable with a perturbative expansion 
\bes
 \obs_s(\mu) = k\,\gbar^2_\msbar(\mu)\,[1 + c_1^s\gbar^2_\msbar(\mu)+\ldots] 
   \label{e:obsexp}
\ees
and define the coupling in the associated physical scheme via
\begin{eqnarray}
  \gbar^2_s(\mu) \equiv \obs_s(\mu) / k = 
   \gbar^2_\msbar(\mu) + c_1^s\gbar^4_\msbar(\mu) +\ldots\,.
   \label{e:gengdef}
\end{eqnarray}
{\em Short-distance} means that $\obs_s(\mu)$ is defined in terms 
of fields concentrated within a 4-d region of linear size $R=1/\mu$. 
In this way, $\mu$ is the only energy scale that enters and the coupling 
runs with $\mu$. {\em Observable} simply means that all $c_i^s$ (or more precisely
$\obs_s$ itself) are finite; no renormalization beyond the one
of the coupling and quark masses is needed. 

While it is not easy to start from experimentally accessible cross sections and directly relate them to such quantities,
sufficient inclusiveness / smearing over energy makes it possible to
approximately define physical couplings in terms of experimental data. However, a direct relation to experimental numbers is not really necessary, rather 
it is sufficient that the same theory and bare coupling uniquely 
predict the physical coupling and experimental quantities such as 
the mass of the proton or decay constants of pion and kaon,
$\fpi,\fK$. For a lattice computation this means that there is a
great opportunity to choose coupling definitions which can be handled
well technically. We will choose two different (families of) couplings,
for reasons to be mentioned below. Note also that 
-- rather exceptionally -- it is a
true advantage that lattice gauge theory works in Euclidean space-time.

We turn to the $\Lambda$-parameters. In massless renormalization
schemes, which we assume throughout,\footnote{The issue of effective
theories with different $\nf$ and quark mass thresholds will be discussed
below.} the integration of the Callan--Symanzik equation 
\bes 
 \mu \partial_\mu \gbar_s(\mu)  = \beta_s(\gbar(\mu))  
\ees
yields the exact relation (at any $\mu$)
\begin{eqnarray}
    \Lambda^{}_s &=& \varphi^{}_s(\gbar^{}_s(\mu))\, \times\, \mu\,,
    \label{e:Lam} 
\end{eqnarray}
with
\begin{eqnarray}
    \varphi_s(\gbar_s) &=& ( b_0 \gbar_s^2 )^{-b_1/(2b_0^2)} 
        \rme^{-1/(2b_0 \gbar_s^2)} \label{e:phig}  \\
     & \times& \exp\bigg\{-\int\limits_0^{\gbar_s} \rmd x\ 
        \Big[\frac{1}{\beta_s(x)} 
             +\frac{1}{b_0x^3} - \frac{b_1}{b_0^2x} \Big] \bigg\} \,. 
        \nonumber                                
\end{eqnarray}
The parameters $\Lambda_s$ are renormalization group invariant,
i.e. independent of $\mu$ and, together with $\beta_s$, give the 
coupling at any $\mu$. Starting from the above equations it is a 
simple exercise to derive the {\em exact} relation of $\Lambda$-parameters
$\Lambda_s/\Lambda_\msbar= \exp(c_1^s /(2b_0))$. Here one-loop PT yields 
the non-perturbative result. This is one reason, why we aim for the 
$\Lambda$-parameter. The second is that once we thus have converted 
from our scheme to
$\Lambda_\msbar$, the coupling can be computed by inserting the 
perturbative approximation 
$$
\betapert_s(g)=-g^3b_0 -g^5b_1 -g^3 \sum_{n=2}^{\lb-1} b_{n,s} g^{2n}
$$
into \eq{e:Lam} and \eq{e:phig}
for $s=\msbar$. There, $\lb=5$ loops are known \cite{MS:4loop1,Czakon:2004bu,Baikov:2016tgj}, making the correction term 
\bes
\Delta \Lambda_s/\Lambda_s = \Delta \varphi_s/\varphi_s = 
c_\lb\alpha^{\lb-1}(\mu) + \ldots
\label{e:perterr}
\ees
due to the difference $\betapert_s-\beta_s$ very small 
{\em in the region where PT applies at all}.\footnote{
The coefficients
$c_\lb$, are, for
$\lb\leq 5$, of order one in the $\msbar$ scheme and expected to be
so in ``good'' schemes in general.}

As we will see, the strategy of the ALPHA collaboration allows to reach
 $\mu =\rmO(100\, \GeV)$ non-perturbatively and
only there uses PT. The perturbative error 
\eq{e:perterr} is then around $[\alpha_\mathrm{SF}(100\, \GeV)]^2 = 10^{-2}$ since
in the used SF scheme, the function 
 $\betapert_\mathrm{SF}$ is known to $\lb=3$ loops.

\section{Non-perturbative $\alpha_s$: meeting the challenge }
\label{s:chall}

\subsection{Challenge}

As said, we have great freedom in our choice for $\obs_s(\mu)$, defining the coupling,
but it is a challenge to reach large $\mu$ (small error term \eq{e:perterr})
in a lattice computation. 
The reason is that numerical computations involve both a 
discretization length, the lattice spacing, $a$, 
and a total size of the system, $L$ that is simulated. For standard 
observables, e.g. the potential of static quarks at short distance, 
there are finite $L$ effects of order $\exp(-\mpi L)$ requiring $L$ to
be several fm. At the same time, one needs to suppress 
discretization errors of order $a^2\mu^2$ and should extrapolate to 
$a^2\mu^2 \to 0$. The resulting inequalities
\bes
  L \gg 1/\mpi\,,\quad 1/a \gg \mu \quad \to \; L/a \ggg \mu/\mpi
  \label{e:chall}
\ees 
lead to the need of very large lattices. To get a feeling for numbers, 
we show a semiquantitative plot of the region
in $\alpha(\mu)$ which enters \eq{e:perterr}
vs. $a^2\mu^2$, which determines the size of the (minimal) discretization
errors for the range of lattice spacings $a > 0.04\, \fm$
reached in the simulations that dominate the present estimates
of $\alpha_s$ by the PDG and FLAG \cite{Agashe:2014kda,Aoki:2016frl}. The 
desired $(0,0)$ point in that plot can only be reached 
by large extrapolations.
\begin{figure}[t]
\hspace*{-7mm}\includegraphics*[width=1.18\linewidth]{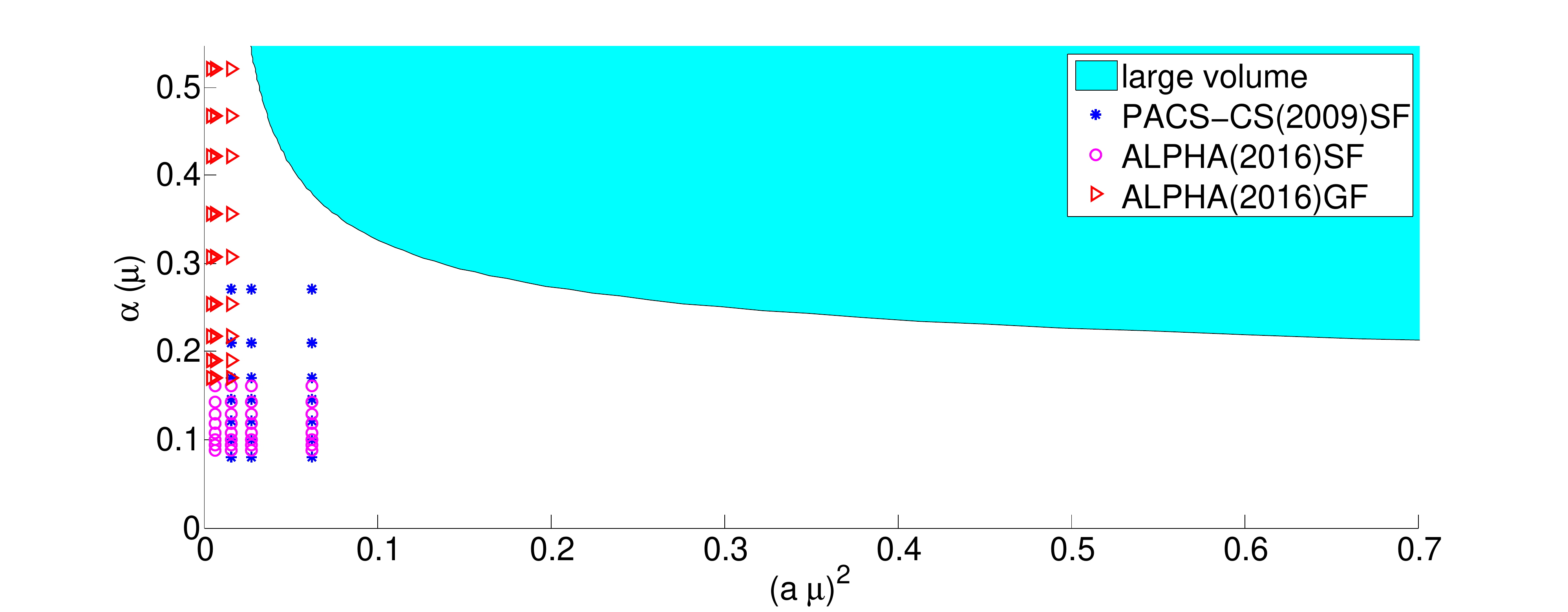} 
  \caption{\label{f:scetch}
  The shaded area shows the region of lattice spacings 
  $a > 0.04\fm$ of present day large volume simulations with
  $\alpha_\msbar(\mu)$ evaluated to two loops and $\Lambda^{(3)}=332\,\MeV$.
  The data points on the left are simulation points in 
  the finite size scaling computations.
}
\end{figure}

\subsection{Finite size schemes}

However, it has been proposed long ago~\cite{Luscher:1991wu}, that one may identify the scales $R=L=1/\mu$ by introducing an
observable $\obs_s$ which depends only on the scale $L$, not on any other ones.
Finite size effects are part of the observable rather than
one of its errors. Instead of \eq{e:chall}
the only restriction is 
\bes
  L/a\gg 1 \,,
\ees
such that $L/a=10 - 50$ lattices are sufficient. Apart from the 
definition of such observables, it remains
to clarify how one connects the perturbative region (large $\mu$ where one can use \eq{e:Lam}, \eq{e:phig} with perturbative $\beta_s=\beta_s^\mathrm{pert}$) with the hadronic region (large $L$, where $\exp(-\mpi L)$ effects
are negligible and one can connect the theory parameters to 
$\fpi,\fK$). This is achieved by

\subsection{Step scaling}

\begin{figure}[t!]
  \centering
  \includegraphics[width=0.6\linewidth]{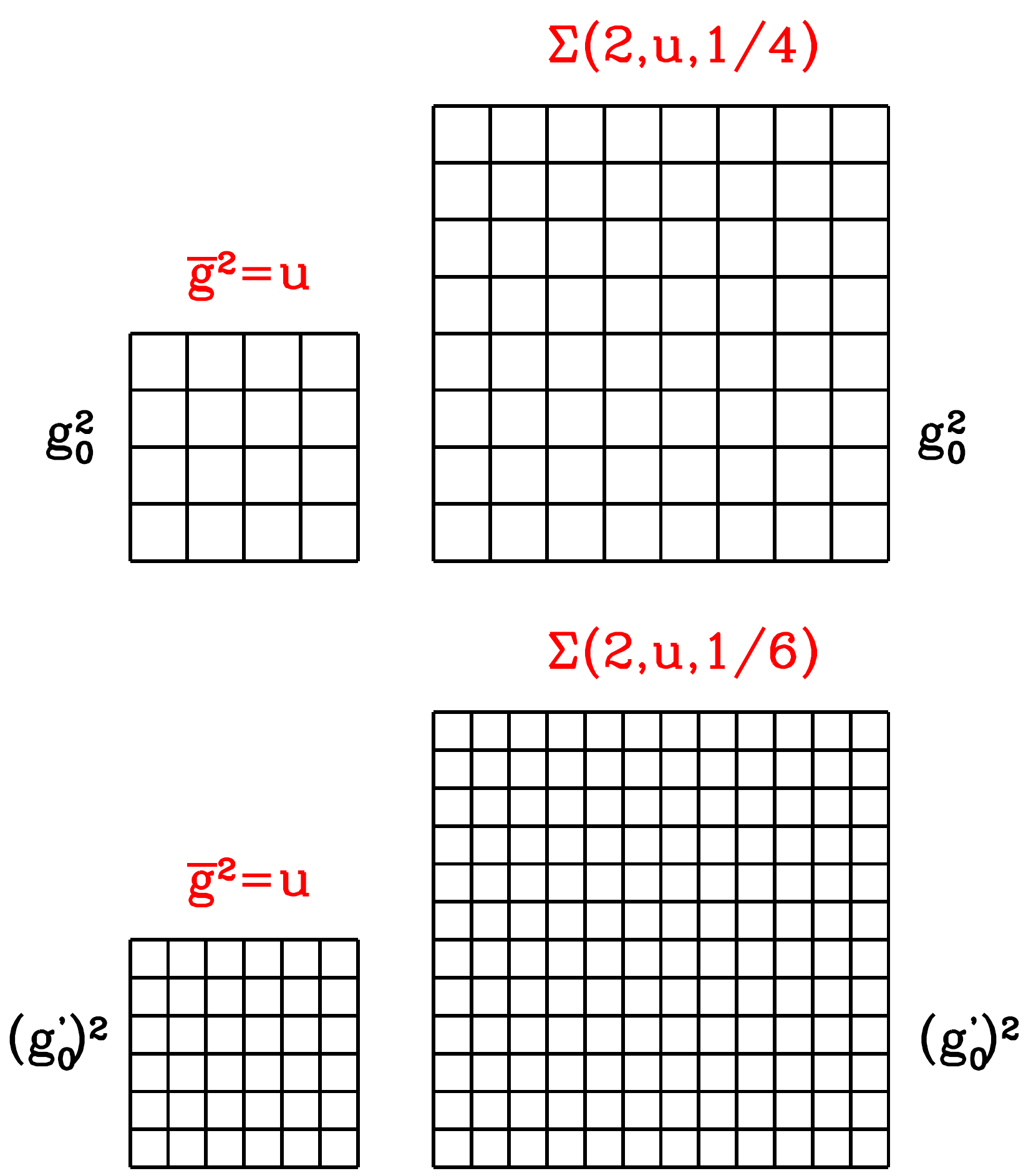}
  \caption{Illustration of the computation of 
  the continuum step scaling function from finer
  and finer lattices. $\Sigma(2,u,a/L)$ in the 
  illustration corresponds to our lattice step scaling function 
  $\Sigma(u,a/L)$. The two different lattice spacings 
  mean two different values $g_0^2$ and $(g_0')^2$ of the bare coupling.
}
\label{f:latt_ssf}
\end{figure}

One replaces the derivative of the coupling with respect to the scale, i.e. the $\beta$-function, by the change of the coupling 
when the scale is varied by a factor of two~\cite{Luscher:1991wu}.
\begin{eqnarray}
  \sigma(u) \equiv
  \left. \gbar^2(\tfrac1{2L})
  \right|_{\gbar^2(\frac1L)=u,m=0} = u + 2b_0 \log(2)\ u^2+ \ldots\,
  \nonumber
\end{eqnarray}
is called the \SSF. Non-perturbatively it 
is computed as the continuum limit 
\begin{eqnarray}
  \sigma(u) = \lim_{a/L\to 0} \Sigma(u,a/L)\,
\end{eqnarray}
of its lattice approximants $\Sigma$.
At finite lattice spacing
the conditions $\gbar^2(1/L)=u$ and $m=0$ refer
to a $(L/a)^4$ lattice and fix the  bare coupling and bare quark mass of the theory. $\gbar^2(1/(2L))$ is 
evaluated for the same bare parameters on a $(2L/a)^4$ lattice, cf. figure~\ref{f:latt_ssf}.
Setting
$m=0$ ensures the quark mass independence of the scheme~\cite{Weinberg:1951ss}. A recursion
\begin{eqnarray}
  u_k = \sigma(u_{k+1}),
\end{eqnarray}
then provides us with $\gbar^2$ at discrete points along the energy axis,
\begin{eqnarray}
  \gbar^2(\mu_k=2^{k} / L_0) = u_k\,, \; k=1,2,\ldots\,.
\end{eqnarray}
Ten such steps cover three orders of magnitude in $\mu$. 

\begin{figure}[ht!]
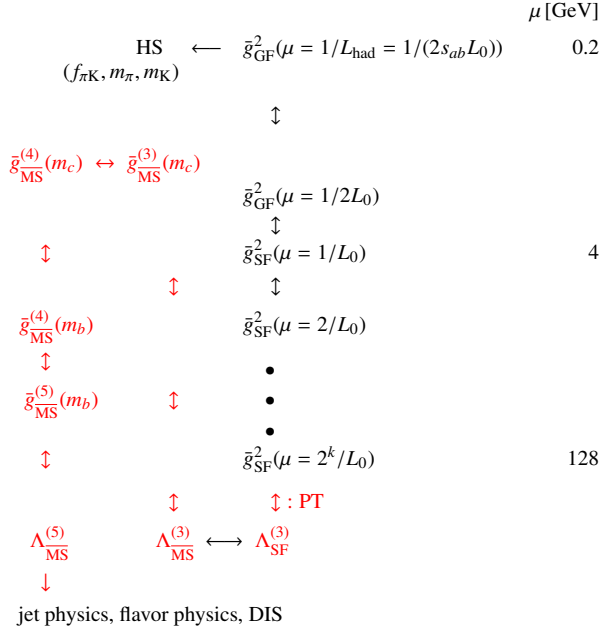

{\footnotesize 
\begin{align*}
                                                       &                                               &  \mu\, [\GeV]
\\
  {\rm HS} \quad \longleftarrow \quad                  &\gbarGF^2 (\mu=1/\Lhad=1/(2\sGF \Lswi))         & 0.2 
\\[-1ex]
  (\fpik,\mpi,\mK)  \qquad\quad 
\\[1ex]     
                                                       &\quad \updownarrow 
\\[1ex]
  \cred \gbar^{(4)}_{\msbar}(m_c)\;\leftrightarrow\;\gbar^{(3)}_{\msbar}(m_c) \qquad 
\\[-1ex]  
                                                    &\gbarGF^2 (\mu=1/{2 \Lswi})  
\\[-0.5ex]
      &\quad \updownarrow&   
\\[-1ex]
 \cred  \updownarrow  \qquad\quad\qquad\qquad\qquad  &\gbarSF^2 (\mu=1/{ \Lswi})                        & 4 
\\
  \cred  \updownarrow \qquad\quad                    &\quad \updownarrow &   
\\
  \cred  \gbar^{(4)}_{\msbar}(m_b) \qquad\quad\qquad\qquad    &\gbarSF^2 (\mu=2/{ \Lswi})&   
\\[-0.5ex] 
  \cred  \updownarrow \qquad\quad\qquad\qquad\qquad  
\\[-2.5ex]                                              &\quad \bullet    
\\[-1ex]
  \cred  \gbar^{(5)}_{\msbar}(m_b) \qquad\quad         \cred  \updownarrow \qquad\quad             &\quad \bullet    
\\[-1ex]
                                                     &\quad \bullet &   
\\[-1ex]
  \cred  \updownarrow \qquad\quad\qquad\qquad\qquad  &\gbarSF^2 (\mu=2^k/{\Lswi})                      &  128
\\[1ex]
  \cred\updownarrow\quad\qquad  &\quad  \cred \updownarrow {\rm \footnotesize :PT}  &  
\\ 
  \cred \Lambda^{(5)}_\mathrm{\msbar}
  \qquad\qquad\cred \Lambda^{(3)}_\mathrm{\msbar} \;\, {\cbla\longleftrightarrow}  &\cred \;\;\Lambda^{(3)}_\mathrm{SF}  
\\ 
 \cred  \downarrow \qquad\quad\qquad\qquad\qquad      \\
\mbox{jet physics, flavor physics,} & \mbox{ DIS} 
\end{align*}
}
\vspace{-0.4cm}
\caption{Our strategy for the computation of 
         the three-flavor $\Lambda^{(3)}$
         ,
         followed by the exact translation to the $\msbar$-scheme 
         (right half). 
         \newline
         On the left we sketch the perturbative connection of 
         the 3-flavor effective theory to the 5-flavor effective 
         theory by standard matching relations at the quark thresholds. 
         HS refers to the hadronic scheme used at low energy. Red parts 
         involve PT with corresponding uncertainties.
\label{f:strategy}}
\end{figure}

In the 90's and 00's a suitable definition of the coupling
was developed~\cite{Luscher:1992an,Luscher:1993gh} and
the above programme was carried out for $\nf=0,2$
by the ALPHA collaboration.
PACS-CS applied the same strategy for $\nf=3$~\cite{Aoki:2009tf}
and partial results are available for
$\nf=4$~\cite{Tekin:2010mm}. The review \cite{Sommer:2015kza} contains more references.
In the following, we 
report on our new results for $\nf=3$ which achieve
a precision which far exceeds previous ones and leads to
a determination of $\alpha_\msbar(m_Z)$ as precise as
the current world average and -- as we would argue -- with much 
improved systematic control over perturbative errors and 
discretization effects.

\section{An optimized strategy}
\label{s:strat}

One reason for the enhanced precision is that 
at approximately $4\,\GeV$ we switch 
to a new scheme \cite{Fritzsch:2013je,DallaBrida:2016kgh}
which has much better statistical accuracy for small $\mu$.
Unfortunately, the computation now has an increased number of
steps to be explained. Here we can just give 
an overview following the sketch in \fig{f:strategy}. 

\begin{figure}[t!]
   \includegraphics*[width=\linewidth]{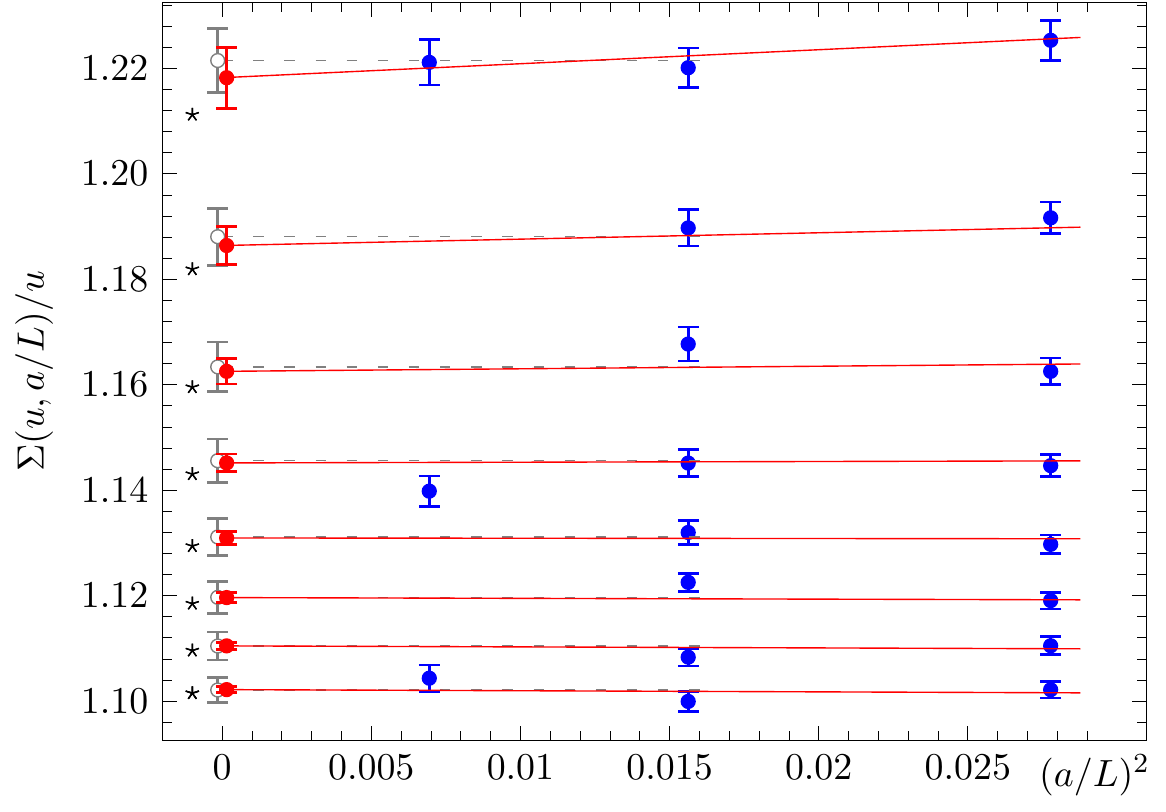}
  \caption{\label{f:cont}Continuum limit of step scaling function in the 
  SF scheme after subtraction of cutoff effects to $2$ loops \protect\cite{Brida:2016flw}. The $\star$-symbols 
   show the perturbative $\sigma$ computed from the three-loop 
   $\betapert$.
   \label{f:SigSF}}
\end{figure}

\section{High energy region: \SF\ coupling}
\label{s:SF}
{\
Our scheme~\cite{Luscher:1993gh} for the high energy 
region, reviewed in ~\cite{Luscher:1998pe,Sommer:2006sj,Sommer:2015kza},
is based on the so-called
Schr\"odinger functional (SF) \cite{Luscher:1992an,Sint:1993un}. 
Here, we just summarize what
is needed to judge our findings below.
Dirichlet boundary conditions are imposed in Euclidean time,
\begin{eqnarray}
    A_k(x)|_{x_0=0} = C_k\,,
    \;\;
    A_k(x)|_{x_0=L} = C_k'\,, \;   k=1,2,3,
\end{eqnarray}
and the gauge potentials $A_\mu$ are taken periodic in space with period $L$.
The six dimensionless matrices 
\begin{align}
LC_k &= i \,{\rm diag}\big( \eta-\tfrac{\pi}{3}, \eta(\nu-\tfrac{1}{2}), -\eta(\nu+\tfrac{1}{2}) +\tfrac{\pi}{3} \big) \,,
\nonumber \\
LC^\prime_k &= i \,{\rm diag}\big( -(\eta+\pi), \eta(\nu+\tfrac{1}{2}) +\tfrac{\pi}{3},-\eta(\nu-\tfrac{1}{2})+\tfrac{2\pi}{3} \big)\,,
\nonumber
\end{align}
depend on the two real parameters $\eta,\nu$.

With these boundary conditions the field which minimizes the action is
unique up to gauge equivalence~\cite{Luscher:1993gh} and denoted by $A_\mu = B_\mu^{\rm class}$. It is a 
constant Abelian color electric field, given in the temporal gauge, $B_0=0$, by
$B_k^\mathrm{class}(x) = C_k + (C_k'-C_k)x_0/L $.  
A family of couplings~\cite{Sint:2012ae}, $\bar{g}_\nu$, is then obtained by
taking $1/\obs_\nu$ in \eq{e:gengdef} to be the $\eta$-derivative of the effective action.}
This yields a simple path integral expectation value,
\begin{eqnarray}
  \langle \partial_\eta S \rangle_{\eta=0} = \frac{12\pi}{\gbar^2_\nu}\,,
\end{eqnarray}
well suited for a Monte Carlo evaluation in the 
latticised theory. Small fluctuations around the background field
generate the non-trivial orders in PT.
The whole one-parameter family of couplings
can be obtained from numerical simulations at $\nu=0$,
since we have
\begin{equation}
  \label{eq:gbarnu}
   \frac{1}{\gbar^2_\nu}  =  \frac{1}{\gbar^2} - \nu\, \vbar\,,
\end{equation}
with $\gbar^2 \equiv \gbar^2_{\nu=0}$ and 
$12\pi\vbar=-\langle \partial_\nu\partial_\eta S \rangle_{\eta=\nu=0}$.

\begin{figure}[t!]
  \centering
  \includegraphics*[width=0.95\linewidth]{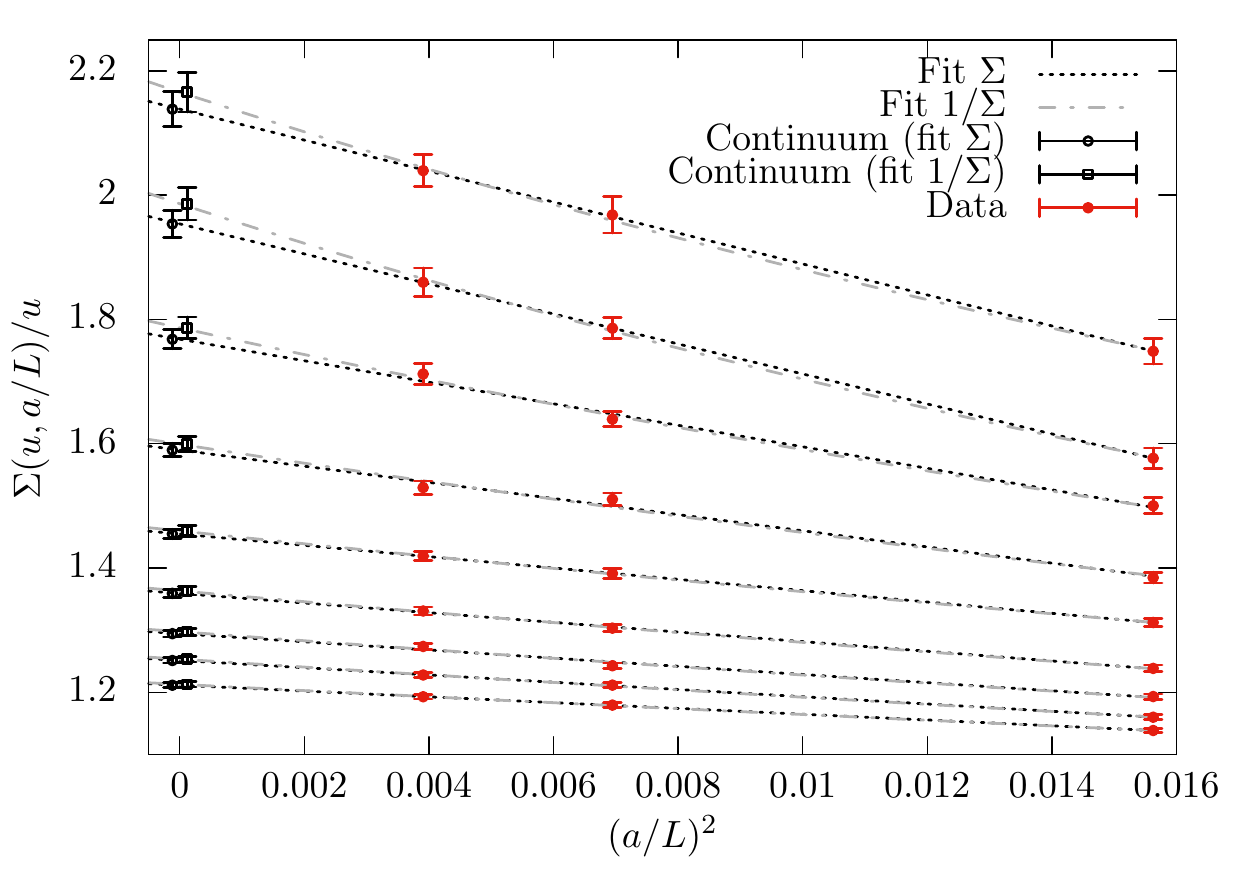}
  \caption{Continuum extrapolation of $\Sigma$ in our GF scheme \protect\cite{DallaBrida:2016kgh} using the discretized flow of 
  \protect\cite{Ramos:2015baa}.}
  \label{f:SigGF}
\end{figure}

Advantageous properties of these couplings are:
1. at large $\mu$ the statistical error decreases 
    proportional to $\gbar^4$.
2. The typical $\sim \mu^{-1},\mu^{-2}$ renormalon contributions
    are absent due the finite volume  infrared momentum cutoff. 
    Instead, the leading known non-perturbative contribution is of order
    $(\Lambda/\mu)^{3.8}$~\cite{Brida:2016flw}.
3.  The $\beta$-function is known including its three-loop term. It is
    well behaved.
4.  As shown in \fig{f:SigSF} and discussed in \cite{Brida:2016flw}    
    discretisation effects are very small. 
    At tree-level of perturbation theory they are $\rmO((a/L)^4)$. We
    subtract~\cite{deDivitiis:1994yz} the known perturbative pieces including two-loop order~\cite{Bode:1999sm}.

\begin{figure}[t]
   \includegraphics*[width=0.9\linewidth]{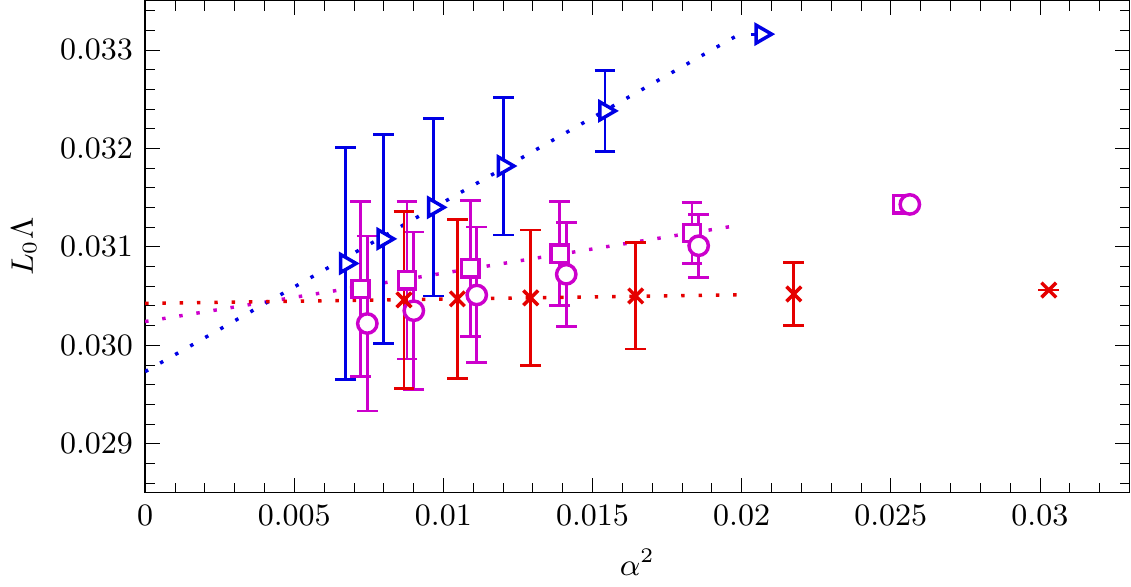}  
  \caption{\label{f:LLmax-extrap} 
   The dependence of the $\Lambda$-parameter on
   the coupling, $\alpha$.  
   From right to left, $k=0,1, \ldots,5$ steps of non-perturbative step-scaling are 
   performed to arrive at $\alpha(\mu)$ at $\mu=2^k/L_0$, before 
   using perturbative running. From top to bottom the different symbols 
  correspond to $\nu=-0.5,0,0.3$. For $\nu=0$ two different ways
  of performing the continuum limit are shown. Dotted straight lines guide the eye.
  \label{f:llplot}}
\end{figure}

The main downside of the SF scheme (see \cite{Brida:2016flw} for details)
is that at larger couplings 
the precision deteriorates. This is avoided by the switch to the GF scheme.
\\
{\bf Results:}
We performed a careful tuning of the bare parameters to have $m\approx0$ within
sharp limits
 and to have 8 fixed values of $\gbar^2(1/L)=u$ on 
$L/a=4,6,8,12$ lattices. We computed $\Sigma(u,a/L)=\gbar^2(1/2L)$ and extrapolated to the continuum limit as sketched in \fig{f:SigSF}. Also $\vbar(L),\vbar(2L)$ were computed and the function
$\omega(u)=\lim_{a/L\to 0} \left.\vbar(L)\right|_{\gbar(L)=u}$ was obtained. 
These results allow to start at $\gbar^2_\nu(1/L_0)$ where 
\bes
  \gbar^2_\mathrm{SF}(1/L_0)=2.012 \quad \mbox{at } \nu=0
\ees
defines $L_0$, and construct the coupling at $\mu=2^k/L_0$ for $k\leq 5$ and for different
$\nu$. We can then compute effective values for $L_0\Lambda_\mathrm{SF,\nu}$ 
using the 3-loop $\beta^\mathrm{pert}$.
These are changed to the $\nu=0$ default scheme via
$L_0\Lambda_\mathrm{SF} = 
\frac{\Lambda_\mathrm{SF,0}}{\Lambda_\mathrm{SF,\nu}}\, L_0\Lambda_\mathrm{SF,\nu} $ with the (exact) ratio $\frac{\Lambda_\mathrm{SF,0}}{\Lambda_\mathrm{SF,\nu}}$. These numbers (points with error bars in \fig{f:llplot})
have to converge to the true $L_0\Lambda_\mathrm{SF}$
with a rate proportional to $\alpha^2(2^k/L_0)$. The numerical
results strongly support this. Given that all data 
points in the graph use PT for $\alpha<0.2$ only, the magnitude 
of differences at finite $\alpha$ is surprisingly big.
In order to exclude that this is a statistical fluctuation, 
we show the function $\omega$ in \fig{f:omega}. It differs by many standard deviations from the two-loop (linear) function, shown in the graph.

These {\bf inaccuracies of PT} do not pose a problem to us 
because we can (and do) simply go to $\alpha\lesssim 0.1$ 
but they are a warning about estimating uncertainties of perturbative
predictions.

\begin{figure}[t]
   \includegraphics*[width=0.9\linewidth]{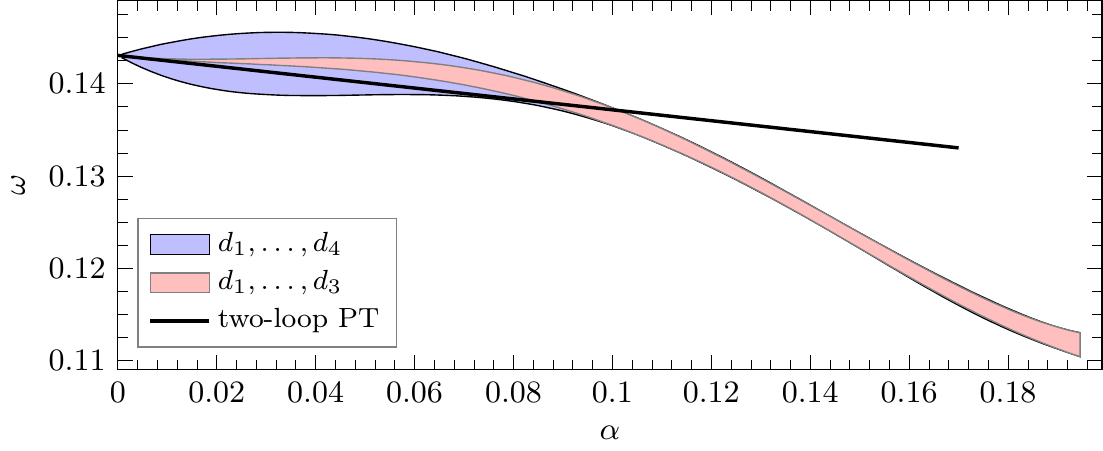}
  \caption{\label{f:omega} The function $\omega(\gbar^2)$ after continuum
  extrapolation, covering the $\pm1\sigma$ band of two fits described 
  in the text. \label{f:omega}
  }
\end{figure}

In \fig{f:strategy} we have now connected $\gbarSF^2 (\mu=1/{ \Lswi})$ to
$\Lambda^{(3)}_\mathrm{SF}$, obtaining
\begin{eqnarray} 
 \Lswi \Lambda^{(3)}_\mathrm{SF} =0.0303(8) \; \,   \to \;
 \Lswi \Lambda_\msbar^{(3)} = 0.0791(21)\,  \label{e:llresult2}
\end{eqnarray} 
and move on to lower energy. 

\section{From 4~GeV to 200~MeV: \GF\ coupling } 
\label{s:GF}
The first step combines the change of scale with the change to 
the new scheme derived from the Gradient Flow 
\cite{Luscher:2010iy,Fritzsch:2013je,Ramos:2015baa,DallaBrida:2016kgh} into
\begin{equation}
  \bar g_{\rm GF}^2(1/2L_0) = 2.6723(64)\,.
  \label{e:phi2}
\end{equation}
The continuum limit is understood. Step scaling functions 
in the GF scheme are then computed and extrapolated to the 
continuum limit in a similar manner as before; only extra care 
has to be taken about higher than $a^2/L^2$ discretization effects
-- a glance at \fig{f:SigGF} shows that $a^2/L^2$ terms are significant. 
We fit the continuum $\sigma$ to
a parameterization of the $\beta$-function,
\begin{eqnarray}
  \beta(g) = \frac{-g^3}{P(g^2) }\,,\quad P(g^2) = p_0 +p_1 g^2 + p_2 g^4 + \ldots\,.
  \nonumber
\end{eqnarray}
using the relation
\begin{eqnarray}
  \label{e:sigmaint}
    \log(2) = 
    -\int_{\sqrt{u}}^{\sqrt{\sigma(u)}} \frac{\rmd x}{\beta(x)} \,.\end{eqnarray}
The parametrization allows us to directly obtain scale factors corresponding to
the change of couplings,
\begin{eqnarray}
  \log(s_{ab}) = \int_{g_a}^{g_b}\rmd x \frac{P(x^2) }{x^3}  \,.
\end{eqnarray}
In particular a careful analysis yields 
\begin{equation}
  s_{ab}= 10.93(20)\quad \mbox{for}\; g_a^2=2.6723\,,
  \; g_b^2= 11.31\,,
  \label{e:sab}
\end{equation}
or combined with \eq{e:phi2} we get $\Lhad/L_0 =  21.86(42)$
where $\gbar^2(1/\Lhad)=11.31$.

From this analysis together with the one in the previous section, 
we also obtained the non-perturbative $\beta$-functions in the two schemes,
in the respective energy ranges considered. A nice graph is found in 
\cite{DallaBrida:2016kgh}.

\section{Hadronic scales}
\label{s:had}
We have to fix $\Lhad$ in physical units from 
$\Lhad = (\Lhad m_\mathrm{had})^{(3)}/m_\mathrm{had}^\mathrm{exp}$
where $m_\mathrm{had}$ is an experimentally accessible low energy mass (scale)
and $(\Lhad m_\mathrm{had})^{(3)}$ is the dimensionless number computed
in QCD with three quark flavors. 
While it is most natural to use the proton
mass, $m_p$, technical limitations explained in detail in
\cite{Sommer:2014mea} lead us to choose the leptonic decay constant 
of pion and kaon, even though their phenomenological values
$\fpi=130.4(2)\,\MeV$ and $\fK=156.2(7)\,\MeV$ depend on the knowledge
of $V_\mathrm{ud}$ and $V_\mathrm{us}$ \cite{Aoki:2016frl}.

Our computation of hadronic scales is based on the
CLS large volume simulations with two degenerate light quarks, $m_u=m_d$ and one additional strange quark~\cite{Bruno:2014jqa}. 
In these simulations the trace, $m_u+m_d+m_s$, of the quark mass matrix 
is held constant~\cite{Bietenholz:2010jr} while varying $m_u=m_d$ in approaching the physical point defined by physical values for
$\mpi/\fpik,\,\mK/\fpik$. 
Along this trajectory in the quark mass plane 
the linear combination
\bes
  \fpik=(2\fK+\fpi)/3\,
\ees 
has a particularly simple dependence on $m_u$. 
Thus it 
can be extrapolated well from the simulation points to the physical point. 
Using this feature, the physical $\fpik$ was related \cite{Bruno:2016plf} to
$t_0^*$, the Gradient Flow scale, $t_0$ introduced by 
M.~L\"uscher \cite{Luscher:2010iy} at the particular 
reference mass point 
\bes
  12 \mpi^2 t_0^* = 1.12 \text{\quad and \quad} m_u=m_d=m_s\,.
\ees
Inserting the phenomenological 
$\fpi$ and $\fK$ yielded 
\bes
   (8t_0^*)^{1/2} = 0.413(5)(1) \,\fm \,.
   \label{e:t0fm}
\ees
Just like our running couplings, it is irrelevant that $t_0^*$
can't be measured directly in experiment. What matters is that
we control the relation to Nature through $\fpik$. 

The scale $t_0^*$
is convenient to finally determine $\Lhad$ in physical units
because it is defined in the mass-degenerate theory with
quark masses far heavier than the physical up and down quark
masses. Thus there are only two parameters
and, since $\mpi$ is 
larger than in Nature, simulations
are easier and finite size effects are smaller.

\begin{figure}[t]
   \hspace*{-6mm}\includegraphics*[width=1.14\linewidth]{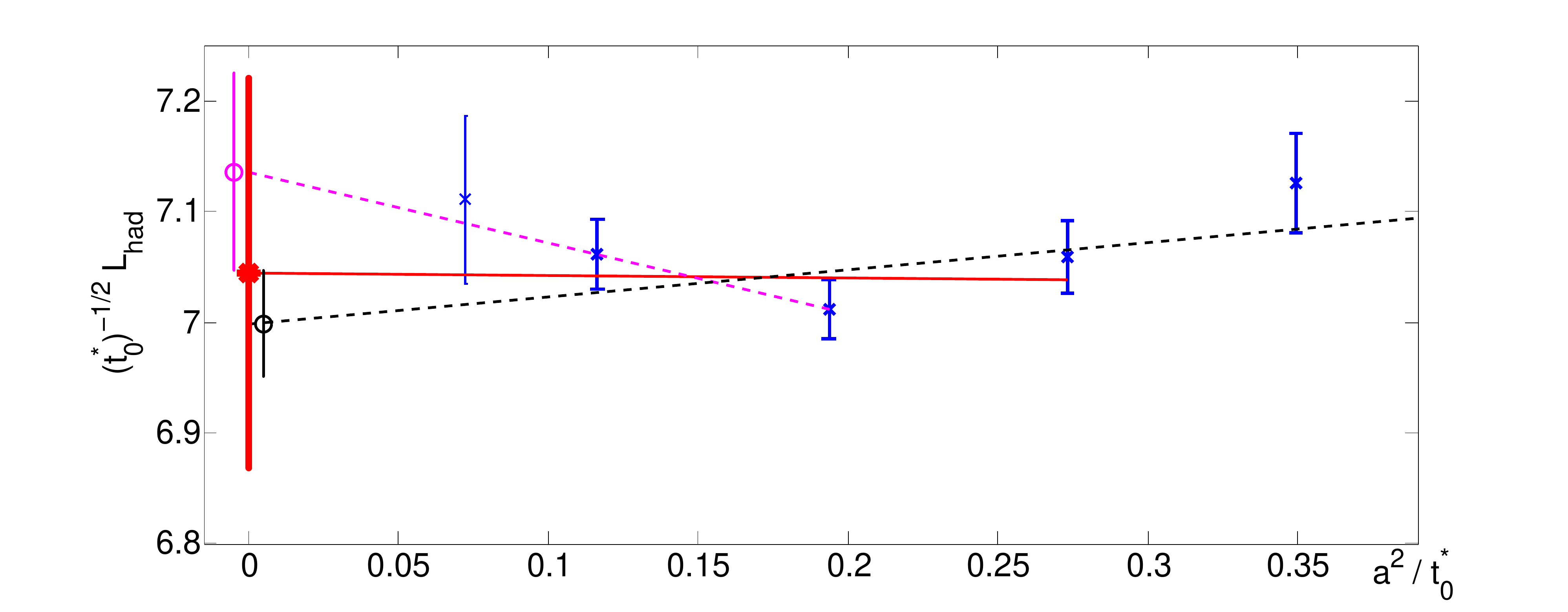}
  \caption{\label{f:lmaxt0} 
  Preliminary continuum extrapolation of  $({t_0^*})^{-1/2}\lmax$.
  The large volume simulation with the smallest lattice spacing is unfinished
   and therfore only included as an illustration. Extrapolations with 
  4,3 and 2 data points are shown together with a range
  for the continuum value covering all of them. 
  It is taken as our preliminary result. 
  }
\end{figure}

These properties enable determinations of $t_0^*/a^2$ and $\Lhad/a$
at five common values of $a$ (or bare coupling $g_0$) 
followed by a continuum extrapolation 
\bes
  ({t_0^*})^{-1/2} \lmax =  \left[({t_0^*})^{-1/2}\lmax \right]_\mathrm{cont} + B\, \frac{a^2}{ t_0^*}
\ees
shown in \fig{f:lmaxt0}. With 
   \eq{e:t0fm}
we then find the preliminary values
\bes
  \lmax= 1.03(3) \,\fm\,, \quad  \Lambda_\msbar^{(3)} = 332(14)\,\MeV\,.
\ees

\section{Connection to the 5-flavor theory and $\alpha_\msbar(m_Z)$}
\label{s:had}
There is little doubt that 3-flavor QCD describes the low energy ($E$)
phenomena including $\Lhad \fpik$ with high precision~\cite{Bruno:2014ufa,Aoki:2016frl}. In other words, the $(E/m_c)^2$ corrections in the effective theory expansion
are small. However,
$\Lambda^{(3)}$ needs to be related to $\Lambda^{(5)}$ because 
physical processes at high energies need 
$\nf\geq5$-flavor QCD and the standard $\alpha_\msbar(m_Z)$ is 
defined in the $\nf=5$ theory.

It has long been known how to connect these theories perturbatively \cite{Weinberg:1980wa,Bernreuther:1981sg} and we now have
4-loop precision \cite{Chetyrkin:2005ia,Schroder:2005hy} in the relation 
\bes
 \gbar^{(\nf-1)} (m_*) = \gbar^{(\nf)} (m_*) (1 + \rmO([\gbar^{(\nf)}(m_*)]^4)\,,
 \label{e:gnfm1}
\ees
where $m_*=\mbar_\msbar(m_*)$ is the mass of the decoupled quark. Together with \eq{e:Lam} and 
$\beta\to\beta^\mathrm{pert}$, we obtain the ratio of the $\Lambda$-parameters. 
We illustrate this by the (red) steps on the left in \fig{f:strategy}.

With the available perturbative precision, we find
\bes
     \Lambda_\msbar^{(5)}&=& 207(11)\,\MeV\,,\; 
      \\[-0.5ex]
    \alpha_\msbar(m_Z)&=& 0.1179(10)(2)\,. 
   \label{e:alphamz}
\ees
The first error in $\alpha$ is just propagated from the one in $\Lambda$, 
which in turn is obtained by standard error propagation of all previously 
discussed numbers which were put together. The second error represents 
our estimate of the uncertainty from using PT in the connection
$\Lambda_\msbar^{(3)}\to\Lambda_\msbar^{(5)}$. We arrive at it as follows.
The $2,3,4$-loop terms in \eq{e:gnfm1} combined
with the  $3,4,5$-loop running lead to contributions
$109,\,15,\,7$
(in units of $10^{-5}$) to  $\alpha_\msbar(m_Z)$. 
We take the sum of the last two contributions 
as our error in \eq{e:alphamz}. {\em Within PT}, this represents a very conservative error estimate: the known terms of the series behave
similar to a convergent series but we treat it like an asymptotic one.
The possibility remains that PT is entirely misleading
when we apply it at $\mu=m_c$, decoupling the charm quark. 
As long as
we do not have a computation of all the above steps with $\nf=4$,
we have to live 
with this -- in our opinion unlikely \cite{Bruno:2014ufa} -- possibility. It would mean that the
second error estimate is far off due to a breakdown of PT for 
 $\Lambda^{(3)}/\Lambda^{(4)}$.

We {\bf thank} the following computer centres and institutions for
computing resources and support: HLRN in Berlin, NIC at DESY, 
Gauss Centre for Supercomputing (GCS) in Munich and J\"ulich , 
Altamira HPC facility at the University of Cantabria, and 
PRACE.
 










\end{document}